\documentclass[final,12pt,authoryear]{elsarticle}

\usepackage{bm}
\usepackage[utf8]{inputenc}

\usepackage[english]{babel}

\usepackage[T1]{fontenc}

\usepackage{lineno}

\usepackage{graphicx}
\usepackage{float}

\usepackage[hidelinks]{hyperref}
\usepackage{xcolor}
\hypersetup{
    colorlinks,
    linkcolor={red!50!black},
    citecolor={blue!50!black},
    urlcolor={blue!80!black}
}

\usepackage{amssymb}
\usepackage{amsmath}        % For aligning

\usepackage{caption}
\usepackage{subfigure}
\usepackage{svg}
\usepackage{media9}
\usepackage{natbib,hyperref}
\newcommand{\mycomment}[1]{}

\journal{\underline{Elsevier}}

\begin{document}
\
\begin{frontmatter}


\title{A micro-mechanics based extension of the GTN continuum model accounting for random void distributions}
%\tnotetext[label1]{This article was presented at the IUTAM Symposium on Size-Effects in Microstructure and Damage Evolution at Technical University of Denmark, 2018}
\author[rvt]{I. Holte}
\author[rvt]{K.L. Nielsen} 
\author[rvtb]{E. Mart\'{\i}nez-Pa\~neda}
\author[rvt]{C.F. Niordson\corref{cor1}} 
\ead[cor1]{cfni@dtu.dk}
\cortext[cor1]{Corresponding author}

\address[rvt]{Department of Civil and Mechanical Engineering, Solid Mechanics, Technical University of Denmark, Kgs. Lyngby, Denmark}
\address[rvtb]{Department of Civil and Environmental Engineering, Imperial College, London SW7 2AZ, UK}
\begin{abstract}
% Abstract
%
Randomness in the void distribution within a ductile metal complicates quantitative modeling of damage following the void growth to coalescence failure process. Though the sequence of micro-mechanisms leading to ductile failure is known from unit cell models, often based on assumptions of a regular distribution of voids, the effect of randomness remains a challenge. In the present work, mesoscale unit cell models, each containing an ensemble of four voids of equal size that are randomly distributed, are used to find statistical effects on the yield surface of the homogenized material. A yield locus is found based on a mean yield surface and a standard deviation of yield points obtained from 15 realizations of the four-void unit cells. It is found that the classical GTN model very closely agrees with the mean of the yield points extracted from the unit cell calculations with random void distributions, while the standard deviation $\textbf{S}$ varies with the imposed stress state. It is shown that the standard deviation is nearly zero for stress triaxialities $T\leq1/3$, while it rapidly increases %in the interval $4/3\lesssim T \lesssim 5$
for triaxialities above $T\approx 1$, reaching maximum values of about $\textbf{S}/\sigma_0\approx0.1$ at $T \approx 4$. At even higher triaxialities it decreases slightly. The results indicate that the dependence of the standard deviation on the stress state follows from variations in the deformation mechanism since a well-correlated variation is found for the volume fraction of the unit cell that deforms plastically at yield. Thus, the random void distribution activates different complex localization mechanisms at high stress triaxialities that differ from the ligament thinning mechanism forming the basis for the classical GTN model. A method for introducing the effect of randomness into the GTN continuum model is presented, and an excellent comparison to the unit cell yield locus is achieved.

\end{abstract}
\end{frontmatter}
%
%
% Introduction
%
\section{Introduction}
\label{S:1}
\noindent{}The loss of load-carrying capacity marks the end of the ductile failure process and occurs either through localization of plastic flow~\citep{tekouglu2015localization,guo2018void,liu2019micromechanism} or by macroscopic homogeneous yielding~\citep{hure2021yield}. The complex sequence of micro-mechanisms controlling failure has been studied for decades~\citep[][]{Tvergaard_1990,Benzerga_Leblond_2010} and it is widely accepted that ductile damage predictions in component-sized structures require continuum modeling for computational efficiency. Thus, numerous homogenized yield criteria have been developed for ductile failure modeling, with the most famous being the model by~\cite{gurson1977}, which explicitly accounts for the porosity, $f$, of the material. However, the Gurson model was early on adjusted by 
\cite{tvergaard1981} and \cite{tvergaard1984analysis} to improve the model accuracy (through the Tvergaard-constants $q_1$ and $q_2$) and account for accelerated void growth at microscopic localization (through the coalescence model imposed through a critical porosity $f^*$). This widely used combined model is nowadays known as the Gurson-Tvergaard-Needleman (GTN) model. However, many other extension has been proposed over the years~\citep[see][and the reference herein]{Benzerga_Leblond_2010}, and while most focuses on obtaining homogenized material properties, the present study will quantify the effect of random void distributions and illustrate how the variation on meso-scale properties can be included in a modified GTN model.

Realistic void distributions are far from periodically arranged, and both Scanning Electron-microscopy and X-ray tomography experiments of voided materials have demonstrated either random or clustered configurations~\citep{Buffiere_et_al_1999,Lecarme_et_al_2014,hannard2017ductilization} - a property not built into the usual continuum-based models. Despite this, the void distribution effects have proven important to the microscopic localization process~\citep{dubensky1987void,magnusen1988effect} and the macroscopic fracture in ductile plate tearing~\citep{Tekoglu_Nielsen_2019,Andersen_et_al_2020b,Celik_et_al_2021}. Generally, a lower ductility is obtained for random distributions due to the triggering of local plastic yielding and loss of load-carrying capacity. For example,~\citet{becker1987effect} performed finite element simulations of a 2D model for an elastic-viscoplastic material obeying the Gurson-Tvergaard yield criterion with an inhomogeneous distribution of voids.  The results showed that plastic flow concentrates into bands in areas with large volume fractions of voids. \citet{perrin1990analytical} analyzed a composite sphere of two porous plastic materials, each obeying the Gurson criterion with individual porosities. The macroscopic yield stress was proven to be different from the one derived from the average homogeneous porosity under hydrostatic stress, indicating that the porosity distribution will affect the effective properties of voided materials.

Unit cell calculations with a single void representing a regular distribution of voids give important fundamental insight into yield properties and damage on a macroscopic scale. However, these material properties are also highly influenced by the spatial void distribution. Many studies have been carried out to investigate details of plastic flow and damage in materials with realistic void distributions. \citet{Fritzen2012} studied statistical effects in large ensembles of voids with volume fractions of up to 30\%, leading to a proposed extension of the Gurson-Tvergaard model in terms of a volume fraction dependence of the correction parameters introduced by \cite{tvergaard1981}. \citet{Khdir2015} demonstrated that the same model could be used for differently shaped voids assuming that the RVE is of sufficient size. Significant dispersion of failure strains, even for large RVEs, was reported through a numerical investigation in \cite{Cadet2021,Cadet2022}. While small RVEs with a single void may not provide a sufficiently realistic representation of macroscopic properties in terms of plasticity and damage, very large RVEs may yield results that do not adequately represent the microscopic variation in material properties on a small scale. On the other hand, intermediate RVEs with a few voids may be used to represent the spatial statistical variations on the scale of individual integration points in a numerical model, thus leading to an appropriate representation of spatially varying mesoscale properties.

This work aims to understand and quantify the effect of the spatial void distributions in terms of the macroscopic yield stress and its dependence on stress triaxiality. To achieve this, three-dimensional representative volume element calculations with periodic boundary conditions containing four spherical voids of equal size distributed randomly are carried out. The voids are embedded in an elastic-perfectly plastic material. Several randomizations are considered for constant initial void volume fraction and stress triaxiality to bring out the statistical variation of the yield locus. In this way, the aim is not to achieve a homogenized response by pursuing a sufficiently large unit cell but rather to understand how void distribution affects the dispersion around a mean yield locus. A subsequent statistical analysis gives input for a proposed yield surface accounting for the dispersion of the material yield point for different void configurations. A modified GTN model is proposed including statistical variations, and the model compares well to the unit cell simulation results. Finally, a procedure for implementing the new yield surface into a large-scale calculation is presented. 

%
%Several representative volume element meshes are generated to obtain three different initial void volume fractions, each having a total of 15 different random void distributions. The imposed stress states are characterised by a fixed value of stress triaxiality under axisymmetric loading. 
%

The paper is organized as follows: Section \ref{S:3} presents the problem formulation in terms of a unit cell with random void distributions, the modeling approach, and the fundamental quantities for the discussion of results. The results from the unit cell study, alongside a comparison to the classical GTN model and a new extension, are presented and discussed in Section \ref{S:5}. The work is concluded in Section \ref{S:6}.

%
% Problem formulation
%
\section{Problem formulation and modeling approach}
\label{S:3}
\noindent{}This work considers a limit load-type analysis of a porous metal with random distributions of discretely modeled microvoids to determine the statistical variation in the yield surface characteristics. Attention is on axisymmetric stress states in the full range of positive stress triaxiality, and the imposed condition is kept constant for each point evaluated on the yield surface (ensuring proportional loading). The matrix is modeled as a rate-independent, perfectly-plastic von Mises material ($J_2$-flow theory), and a small strain finite element formulation is adopted to mimic the limit load of the material. The matrix material is characterized by the parameters: $\sigma_0/E=0.001$ and $\nu=0.3$, where $\sigma_0$ is the yield stress, $E$ is Young’s modulus, and $\nu$ is the Poisson ratio. The unit cell configuration is described in Section \ref{sec:unitcells}, the modeling approach is outlined in Section~\ref{sec:numericalmethod}, while the fundamental quantities in the statistical analysis and for the discussion of results are outlined in Sections~\ref{S:4} and~\ref{sec:PI}.
\subsection{Unit cell configuration}
\label{sec:unitcells}
\noindent{}Figure~\ref{fig:unitcellmodel} shows the cubic unit cell setup, where $a_0$ denotes the side length along the three coordinate axes $x_i$ $(i=1,2,3)$. Each unit cell contains four ($N=4$) spherical voids defined by their center coordinates and initial radius $r_0$. Here, the non-dimensional total porosity $f_0=\frac{4\pi}{3}\frac{Nr_0^3}{V}$ (or unit cell void volume fraction) of the unit cell with volume $V$ determines the void radius, and the results are presented for three different $f_0$-values in Section~\ref{S:5}. The spatial location of the voids is determined using an ad-hoc algorithm implemented by means of the Abaqus2Matlab software~\citep[see][]{AES2017}. The algorithm generates a given number of 3D spheres inside a 3D domain, with the radii and the positions of the spheres being uniformly random. The algorithm, simplified to equally sized voids, involves the following steps:
\begin{itemize}
%\item[(i)]The radii of the spheres are randomly generated and sorted from larger to smaller (here considering only equally sized voids).
%
\item[(i)]A 3D point grid of potential void centers is created within the unit cell. 
\item[(ii)]The order of the grid points are randomly permuted. 
\item[(iii)]The mutual distances between all center points are calculated and a new center point is defined if the distance is smaller than a minimum distance $2r_0+L$, with $r_0$ and $L$ being the void radius and minimum ligament size, respectively.   
%
%\item[(iv)]All points are scanned, and for each point, the distance to the centers of the spheres is estimated. If the distance is higher than the sum of the current sphere radius, a tolerance seed (determining the intervoid ligament distance), and the radii of the existing spheres, then the grid point is considered a sphere center.
%
\end{itemize}

In this way, the algorithm introduces a set of non-overlapping spherical voids, and a representation of three random configurations is shown in Fig. \ref{fig:meshes} in terms of finite element meshes.

For comparison purposes, the present work also considers a regular Face-Centered Configuration (FCC) void distribution loaded along the cubic axes. Figure~\ref{fig:FCC-shifted}a illustrates how the configuration can be modeled considering only four voids when translating the unit cell in the positive $x_1$-, $x_2$- and $x_3$-direction. The corresponding finite element mesh for the FCC unit cell is shown in Fig. \ref{fig:FCC-shifted}b. It is worth noticing that the FCC unit cell response is independent of the translation along the coordinate axes. 
\subsection{Numerical modeling approach}
\label{sec:numericalmethod}
\noindent{}The present work adopts a small strain finite element formulation and exploits the commercial software package~\cite{ABAQUS_2016}. Thus, the model setup cannot account for the softening owing to void growth. Instead, the load-carrying capacity of the unit cell represents the limit load and, thereby, a point on the yield surface when plotted in stress space. Throughout, axisymmetric stress states with $\sigma_2=\sigma_3$ are considered such that the stress state is defined by the stress ratio
\begin{equation}
    \rho=\frac{\sigma_{3}}{\sigma_{1}}=\frac{\sigma_{2}}{\sigma_{1}}
\label{eq:rho}    
\end{equation}
\noindent where $\sigma_{1}$ is the stress along the main loading axis, and $\rho$ is kept constant and prescribed for each individual analysis of a point on the yield surface. Thus, the von Mises equivalent stress $\sigma_e$, and the overall mean stress $\sigma_m$ are
\begin{equation}
\sigma_e= |\sigma_{1}-\sigma_{2} | =\sigma_{1}(1-\rho) \qquad \text{and} \qquad \sigma_m=\frac{\sigma_1+2\sigma_2}{3}=\sigma_1 \frac{1+2\rho}{3}, 
\label{eq:mises_and_mean}    
\end{equation}
while the stress triaxiality $T$ is related to the stress ratio through
\begin{equation}
 T =\frac{1}{3}\left(\frac{1+2\rho}{1-\rho}\right).
\end{equation}
%
%Here, the overall stress components $\sigma_{i}$ are calculated from the volume average of the overall unit cell, such that 
%%
%\begin{equation}
%\sigma_{i}=\frac{1}%{V}\int_V\sigma_{ij}\text{d}V, 
%\end{equation}
%%
%where $V$ is the unit cell volume. 
%
In the finite element calculations, the ratio $\rho$ between the transverse and axial stress components is kept constant using multiple-point constraints (MPCs in Abaqus). This is achieved by introducing an extra set of degrees of freedom to the finite element mesh in terms of three dummy nodes $N_i$ $(i=1,2,3)$ placed outside the finite element mesh as depicted in Fig.~\ref{fig:RVE_PBC}. The dummy nodes are connected with spring elements (SPRING2 in the Abaqus) to three master nodes $M_i$ $(i=1,2,3)$, which are part of the unit cell mesh. In this way, the displacement of the dummy nodes is related to the forces $F_i$ acting on the faces of the unit cell (along its normal) through
\begin{equation}
\label{eq:Fi}
    F_i = k_i(u_i^{N_i}-u_i^{M_i}), \quad i=1,2,3,
\end{equation}
where $k_i$ are the spring element constants. Moreover, the forces $F_i$ are related to the macroscopic stresses through
\begin{equation}
\label{eq:sigma=F/A}
    \sigma_{1}=\frac{F_1}{A_1}, \quad \sigma_{2}=\frac{F_2}{A_2}=\sigma_{3}=\frac{F_3}{A_3},
\end{equation}
where $A_i$ are the surface areas of the unit cell. Thus, combining Eqs.~\eqref{eq:rho}, \eqref{eq:Fi}, and \eqref{eq:sigma=F/A}, and solving 
for the displacement of the dummy nodes (for constant $\rho$) gives
\begin{align}
\label{eq:MPC}
    u_i^{N_j}=u_i^{M_j}+\rho\left(u_1^{N_j}-u_1^{M_j}\right),  \quad \quad i,j = 1,2,3.
\end{align}
Here, $u_1^{N_1}$ is the prescribed displacement in the main loading direction $x_1$, while the remaining displacements of the dummy nodes are calculated in the MPC subroutine.

Furthermore, to ensure the periodicity of the unit cell, a set of linear constraint equations is imposed according to 
%
%\begin{align}
%\label{eq:PBC}
%&\text{Faces:} \quad \quad \quad \quad \quad \quad \quad \quad \quad \quad \quad \text{Edges:} \nonumber \\
%    &\quad u_i^{ABCD}=u_i^{A'B'C'D'}+u_i^C  \nonumber \quad \quad \quad u_i^{AB}=u_i^{D'C'}+u_i^C+u_i^{B'} \nonumber \\
%    &\quad u_i^{A'ABB'}=u_i^{D'DC'C}+u_i^{B'} \nonumber \quad \quad \quad u_i^{AD}=u_i^{B'C'}+u_i^C+u_i^{D'} \nonumber \\
%    &\quad u_i^{A'BDD'}=u_i^{BB'CC'}+u_i^{D'} \nonumber \quad \quad  \quad u_i^{AA'}=u_i^{CC'}+u_i^{B'}+u_i^{D'} \nonumber \\
%&\text{Vertices:} \nonumber \\
%    &\quad u_i^{A'}=u_i^{D'}+u_i^{B'}  \\
%    &\quad u_i^{D}=u_i^{C}+u_i^{D'} \nonumber \\
%    &\quad u_i^{B}=u_i^{C}+u_i^{B'} \nonumber \\
%    &\quad u_i^{A}=u_i^{C}+u_i^{B'}+u_i^{D'} \nonumber
%\end{align}
%
%
%
\begin{align}
\label{eq:PBC}
&\text{Faces:} \quad \quad \quad \quad \quad \quad \quad \quad \quad \quad \quad \text{Edges:} \quad \quad \quad \quad \quad \quad \quad \quad \quad \quad \quad \text{Vertices:} \nonumber \\
    &\quad u_i^{ABCD}=u_i^{A'B'C'D'}+u_i^C  \nonumber \quad \quad \quad u_i^{AB}=u_i^{D'C'}+u_i^C+u_i^{B'} \nonumber \quad  \quad \quad u_i^{A'}=u_i^{D'}+u_i^{B'} \nonumber \\
    &\quad u_i^{A'ABB'}=u_i^{D'DC'C}+u_i^{B'} \nonumber \quad \quad \quad u_i^{AD}=u_i^{B'C'}+u_i^C+u_i^{D'} \nonumber \quad  \quad \quad u_i^{D}=u_i^{C}+u_i^{D'} \nonumber\\
    &\quad u_i^{A'BDD'}=u_i^{BB'CC'}+u_i^{D'} \nonumber \quad \quad \quad u_i^{AA'}=u_i^{CC'}+u_i^{B'}+u_i^{D'} \nonumber \quad \quad \quad u_i^{B}=u_i^{C}+u_i^{B'} \nonumber\\
    &  \quad \quad \quad \quad \quad \quad \quad \quad \quad \quad \quad \quad \quad \quad \quad \quad \quad \quad \quad \quad \quad \quad \quad \quad \quad \quad \quad ~\quad u_i^{A}=u_i^{C}+u_i^{B'}+u_i^{D'} \nonumber \\
\end{align}
for $i=1,2,3$. Here, $u_i$ is the displacement in the $i$th direction ($i=1,2,3$) of the nodes at the faces, edges, and vertices of the unit cell, as shown in Fig. \ref{fig:RVE_PBC}. The nodes at the corners $B'$, $C$, and $D'$ are the master nodes.

%The combination of multiple point constraints and linear constraint equations ensures that the unit cell is periodic while maintaining a given loading state for each simulation increment.  

\subsection{Statistical variation}
\label{S:4}
\noindent{}The random void distributions naturally introduce a statistical variation in the model output. The yield point extracted at the limit load of the unit cell depends on the localization process between voids (or lack thereof) and, thus, on the intervoid distance and the location of the voids. Thus, a mean $\mu$ and standard deviation $\mathbf{S}$ are introduced to characterize the span of yield points obtained for a specific stress state and initial porosity. Note that proportional loading is imposed using a number of constant stress ratios $\rho$, such that all the determined yield points for a given $f_0$ and $\rho$ will be located on a straight line through the origin in the 
$\sigma_e-\sigma_h$ stress space. Let $s_i=\sqrt{\sigma_e^2-\sigma_h^2}$ be the distance from the origin to the yield point of the $i$th unit cell calculation for a specific value of $\rho$. The mean of $n$ unit cell calculations with different random void distributions is then given by
\begin{equation}
\mu = \frac{1}{n}\sum_{i=1}^n s_i, 
\label{eq:meandist}
\end{equation}
and the corresponding standard deviation of the mean distance is
%%
%\begin{equation}
%\mathbf{S} = \sqrt{\frac{1}{n-1}\sum_{i=1}^n\abs{s_i-%\mu}^2}
%\label{eq:STD}
%\end{equation}
%%
%
\begin{equation}
%\mathbf{S} = \sqrt{\frac{1}{n}\sum_{i=1}^n(s_i-\mu)^2}
\mathbf{S} = \sqrt{\frac{1}{n-1}\sum_{i=1}^n(s_i-\mu)^2}
\label{eq:STD}
\end{equation}
%with the standard error of mean being $\mathbf{SEM}=\mathbf{S}/\sqrt{n}$. The standard error of the mean $\mu$ can be interpreted as a measure for the dispersion of sample means around the population mean given by Eq.~\eqref{eq:meandist}.

\subsection{Characterisation of deformation mechanism}
\label{sec:PI}
\noindent{}To discuss the mechanism of plastic deformation leading to the loss of load-carrying capacity of the unit cell, a plastic index $\mathbf{PI}$ is introduced as 
\begin{equation}
\mathbf{PI} = \frac{V_p}{V_m},
\label{eq:PI}
\end{equation}
where $V_m$ is the volume of the matrix material, and $V_p$ is the volume of the unit cell undergoing plastic yielding. % calculated by the following sum over all integration points $n_g$
%%
%\begin{equation}
%    V_p = \sum_{i=1}^{n_g}\sum_{j=1}^{n_g}\sum_{k=1}^{n_g}\beta W_i W_j W_k ||J||.
%\label{eq:Vp}
%\end{equation}
%%Here, $\beta$ is a flag indicating %whether the current Gauss point is yielding ($\beta=1$) or remains elastic ($\beta=0$), and $n_g$ is the total number of Gauss points, $W_i$ are the Gauss point weights, and $||J||$ is the determinant of the Jacobian matrix. 
%
The plastic index provides a way to distinguish between macroscopic yielding and localization of plastic flow~\citep[or homogeneous versus inhomogeneous yielding in the terminology of][]{hure2021yield}. At the limit load, the major part of the unit cell undergoes macroscopic yielding when $\mathbf{PI}\rightarrow 1$, while small values of $\mathbf{PI}$ indicate localization in part of the unit cell volume.

\section{Numerical results and discussion}
\label{S:5}
\noindent{}In the following, yield surfaces are constructed in the von Mises versus mean stress space for three values of the total porosity  $f_0=0.00085$, $0.017$, and $0.034$ to investigate the influence of the void distribution. The results are obtained by imposing stress states corresponding to nine different values of the stress ratio $\rho=-0.5$, $0$, $0.4$, $0.625$, $0.73$, $0.8$, $0.85$, $0.9$, and $0.99$, spanning the range of positive stress triaxiality. The purely hydrostatic state of stress $\rho=1$ is here omitted due to convergence issues. The calculations for each stress state and porosity are repeated using 15 different randomizations of the void distribution to form a statistical basis for the discussion of results. In addition, the FCC distribution is investigated. Thus, a total of 432 combinations of stress state, void distributions, and void volume fraction are considered.

\subsection{Yield surfaces extracted as the unit cell limit-load}
\noindent{}Figure \ref{fig:YS_clusters_FCC_mean} presents the simulated yield points for all combinations of porosity $f_0$, stress triaxiality $T$, and void distribution considered in the present work. The circular markers show results for unit cells with random void distributions, while the square markers are the mean value given by Eq.~\eqref{eq:meandist} for a specific combination of porosity and triaxiality. Here, solid lines indicate the yield surface represented by the mean values. In addition, triangular markers indicate the corresponding results for the unit cell with a regular FCC distribution, while the overlaying dashed lines illustrate the corresponding yield surfaces. Colors distinguish the results for the different initial porosities $f_0$, and the well known delay in yielding is evident for diminishing initial porosity. This feature is also represented in the classical Gurson-Tvergaard-Needleman (GTN) yield surface~\citep[][]{gurson1977,tvergaard1981,tvergaard1984analysis} which is defined by:
\begin{equation}
\label{eq:Gurson}
    \Phi = \frac{\sigma_e^2}{\sigma_0^2}+2q_1f \text{cosh}\left[\frac{3}{2}q_2\frac{\sigma_m}{\sigma_0}\right]-\left(1+(q_1f)^2\right).
\end{equation}
Here, $\sigma_e$ is the von Mises stress, $\sigma_0$ is the matrix material flow stress, $\sigma_m=\sigma_{kk}/3$ is the mean stress, $f$ is the void volume fraction, and $q_1=1.5$ and $q_2=1$ are the Tvergaard-constants~\citep{tvergaard1981}. As both the von Mises equivalent stress and the mean stress enter explicitly into the GTN yield surface, it can readily be represented in the von Mises versus mean stress space.% For this, the matrix flow stress $\sigma_M$ is substituted by the matrix yield stress $\sigma_0$ and the porosity by the value considered $f_0$. %However, the GTN model cannot represent the statistical variation due to the random void distribution in its current form.

Figure~\ref{fig:YS_clusters_FCC_mean} shows a combined dependence on the void distribution and the stress triaxiality. For the lowest triaxiality values, i.e., the results closest to the von Mises stress axis, the effect of the void distribution is negligible, and all yield points practically coincide. This is in line with results presented in \cite{tekouglu2015localization} and \cite{holte2021interaction}, where $T=1$, corresponding to $\rho=0.4$, is found as the limit below which the onset of macroscopic yielding~\citep[homogeneous yielding according to ][]{hure2021yield} co-occurs with void coalescence, i.e., with intervoid localization. However, the dispersion in the yield points amplifies when increasing the stress triaxiality indicating  the activation of different or more complex localization mechanisms that depends on the interaction of voids and, thereby, their spatial distribution. The results for a specific porosity and stress state show yield points over a significant range of $\sigma_e$ and $\sigma_m$ combinations, indicating that the void distribution is an important microstructural property for porous metals at these triaxiality levels. Moreover, for the highest triaxiality ($\rho\approx 1$), i.e. the results closest to the mean stress axis, the spread in yield points is moderate compared to the slightly lower triaxiality levels. This indicates a shift in the localization mechanism, leading to a smaller dependency on the void distribution.

Comparing results from the random void distributions to that of the FCC unit cells, the regular void distribution generally displays more plastic resistance, although it is not a strict upper bound. Evidently, the yield point for the FCC configuration falls below that of some of the random distributions for triaxiality values in the range $T=1$ to $4$, where a large dispersion in the yield points is observed for the random distributions (see Fig.~\ref{fig:YS_clusters_FCC_mean}). At this triaxiality level, the stresses transverse to the main loading axis increase the propensity to alter the localization mode as the intervoid distances vary across the unit cell. In contrast, the regular FCC configuration has the same ligament geometry between all voids and, thus, will not exhibit the same shift in localization mechanism when changing the stress state. However, the mean for the random void distributions (solid lines) consistently gives less plastic resistance than the FCC configuration (dashed lines), independently of the initial porosity. The difference is prominent at moderate to high triaxialities, while the yield surfaces practically coincide at low stress triaxialities.

The dispersion in the yield points for the random void distributions may be quantified by the standard deviation $\mathbf{S}$ of the distance to the origin (see Eq.~\eqref{eq:STD}). Here, results are based on 15 randomizations of each combination of porosity and triaxiality. Figure~\ref{fig:STD} is constructed from the unit cell results to show the yield loci defined by the mean surface plus minus the standard deviation ($\mu\pm\mathbf{S}$) for the three investigated void volume fractions. It is seen that the standard deviation increases along the mean stress axis, i.e., with increasing triaxiality as discussed for the dispersion concerning Fig.~\ref{fig:YS_clusters_FCC_mean}. That is, the values of $\mathbf{S}$ are small for low stress triaxialities approaching zero as $T\rightarrow 0$, while values of $\mathbf{S}$ are the highest in the interval $T=4$ to $5$. Above this triaxiality level, it decreases slightly as the triaxiality goes to infinity ($\rho\rightarrow 1$). The span of the yield loci in Fig.~\ref{fig:STD} is consistent with the span with the yield points quantified out in Fig. \ref{fig:YS_clusters_FCC_mean}. From Fig.~\ref{fig:STD}, it also becomes clear that the dispersion of the yield point at intermediate levels of stress triaxiality and the narrowing as $\rho \rightarrow 1$ is most prominent for high initial void volume fractions. This can be ascribed to the change between localization mechanisms. For high $f_0$-values, the intervoid ligaments carry higher stresses due to the smaller volume fraction of matrix material, making the ligaments more susceptible to the deformation mechanism involving localization.

To quantify the variation in the dispersion of yield points, the standard deviation normalized by the yield stress $\mathbf{S}/\sigma_0$ is shown as a function of $\rho$ in Fig.~\ref{fig:STD-rho}. Values obtained from the unit cell calculations are circular markers, while a cubic Hermite interpolation of the results is a continuous dashed line.
%\footnote{Only the unit cell values for $\mathbf{S}/\sigma_0$ are provided for the interpolation, and the derivatives in each point are therefore estimated.}.
The results confirm the increase in standard deviation with increasing $\rho$ until a peak is reached around $\rho=0.8$ (corresponding to $T=4.3$), after which the standard deviation decreases at higher triaxiality levels. The peak value and the subsequent drop in the standard deviation are largest for the two highest initial porosities ($f_0=0.017$ and $0.034$), while the decrease at high triaxialities is more modest for the lowest initial porosity ($f_0=0.0085$). The variation in $\mathbf{S}/\sigma_0$ with the prescribed stress state $\rho$ is clearly reflected in the scatter of the results in Fig.~\ref{fig:YS_clusters_FCC_mean}. Moreover, it is noticed from Fig.~\ref{fig:STD-rho} that increasing $f_0$ leads to a larger value for $\mathbf{S}/\sigma_0$ for all triaxiality values.

\subsection{Mechanisms leading to the dispersion of yield points}
\noindent{}The plastic index $\textbf{PI}$ introduced in Eq.~\eqref{eq:PI} is considered in an attempt to link the dispersion of the yield points to the localization mechanism at play in unit cells with random void distributions. The plastic index is calculated at the limit load for all configurations of $f_0$ and $T$ and displayed with circular markers in Fig.~\ref{fig:PI}. The mean value of the plastic index $\mu_{\mathbf{PI}}$ for each porosity value is shown as a function of $\rho$ (solid lines). A plastic index of $1$ corresponds to yielding in the entire unit cell, while low $\textbf{PI}$-values signal intense localization in a smaller part of the unit cell volume.

It is observed from Fig.~\ref{fig:PI} that almost the entire unit cell volume deforms plastically for $\rho\lesssim 0.5$ (corresponding to $T\lesssim 4/3$) for all values of $f_0$ considered. This is well in line with the fact that macroscopic yielding (or homogeneous yielding) is the dominant deformation mechanism at low to moderate values of stress triaxiality, rendering the effect of the void distribution negligible. In contrast, the plastic index shows a much greater dispersion for higher values of $\rho$, ranging from $0.4$ to $1$. The low values of the plastic index reflect intense localization in a small portion of the unit cell, which is highly controlled by the location of the voids and, thereby, the void distribution. At high triaxiality levels, the limit load can be attained through macroscopic yielding, e.g., for equally distanced voids, and microscopic localization if voids are located in a favorable band. Moreover, the large stress components transverse to the main loading direction increases the likelihood of encountering a favorable voided band. It is this combined effect ofdifferent mechanisms or between different favorable localization modes at high stress triaxialities that makes void distribution essential to the plastic resistance. In this way, it is the localization mechanism that controls the dispersion in both the plastic index $\textbf{PI}$ and the yield loci shown in Figs.~\ref{fig:YS_clusters_FCC_mean} and~\ref{fig:STD}. This is also evident from Fig.~\ref{fig:std_PI}, showing the standard deviation of the plastic index $\textbf{S}_\textbf{PI}$ as a function of $\rho$ for all values of $f_0$ considered. For increasing triaxiality, i.e., increasing $\rho$, the standard deviation increases, indicating a greater variation in the plastic deformation mechanism, and the variation compares well to that of $\textbf{S}/\sigma_0$ in Fig.~\ref{fig:STD-rho}. As seen in Fig.~\ref{fig:STD-rho}, the standard deviation for the plastic index $\textbf{S}_\textbf{PI}$ also increases with increased $f_0$ since the size and position of the intervoid ligaments affect the deformation mechanism to a greater extent when the void volume increases.

\subsection{Introducing the effect of randomness into the GTN yield surface}
\label{sec:Gurson_enriched}
\noindent{}Figure~\ref{fig:mean_Gurson} shows the mean distance to the origin $\mu$ based on Eq.~\eqref{eq:meandist} as obtained from the unit cell investigations of random void distributions where each data point and corresponding standard error is based on 15 different randomizations of the void distributions. The small error bars in Fig. \ref{fig:mean_Gurson} show the standard error of the mean, indicating that % the dispersion of the plastic resistance of the sample (the 15 unit cell calculations) is close to the mean distance of the population (any random distribution of the four voids).
a sufficient sample size is used.
%The results show that the mean distance from the origin is smallest at zero triaxiality, while the largest distance exists for $\rho\rightarrow 1$. 
Moreover, the initial void volume fraction has little effect on the plastic resistance distance for $T<1$, while the influence of porosity increases with triaxiality. Comparing the results to the predictions by the classical GTN model, a good agreement is obtained for all values of the initial porosity $f_0$ and stress states $\rho$. Thus, the GTN model may form a basis for a micro-mechanics based continuum model accounting for random void distributions if modified suitably. 

The classical GTN model cannot account for the dispersion of the yield points observed in Fig.~\ref{fig:YS_clusters_FCC_mean} and mapped out in Fig.~\ref{fig:STD-rho}, which are consequences of randomness in the void distribution. However, as demonstrated in Fig.~\ref{fig:STD}, the unit cell response is accurately represented by overlaying the mean distance $\mu$ by the standard deviation $\textbf{S}$ suggesting that the yield locus may be expressed on the form
\begin{equation}
\Phi=\Phi_\mu\pm\Phi_\textbf{S}
\end{equation}
where $\Phi_\mu$ is the mean of the yield locus and $\Phi_\textbf{S}$ represents the spread of the yield surface.
%$\Phi=\mu\pm\textbf{S}$.%
Thus, since the GTN model quite accurately models the mean yield surface (see Fig.~\ref{fig:mean_Gurson}), it is suggested to scale the distance to the origin of the GTN yield surface with the standard deviation such that the distribution-enriched GTN yield surface is expressed as
\begin{equation}
\label{eq:GursonS}
    \Phi = \frac{\sigma_e^2}{\sigma_0^2}+2q_1f \text{cosh}\left[\frac{3}{2}q_2\frac{\sigma_m}{\sigma_0}\right]-\left(1\pm \frac{\mathbf{S}}{\sigma_0}\right)^2\left(1+(q_1f)^2\right),
\end{equation}
Here, $\mathbf{S}$ is the standard deviation which is a function of the porosity $f$, and the stress state $\rho$ according to Fig.~\ref{fig:STD-rho}. A continuous yield function is obtained for the stress states and porosity values considered by incorporating the cubic Hermite interpolation, and Fig.~\ref{fig:GursonS} depicts the new yield surfaces alongside the classical GTN yield surface. The GTN yield surface (dashed lines) represents the mean surface of the unit cell with random void distributions, while the dispersion of yield points is obtained through the scale factor $(1\pm\textbf{S}/\sigma_0)^2$ shown as solid lines on either side of the mean yield surface. The depicted confidence interval of $\pm\textbf{S}/\sigma_0$ encloses $70\%$ of the expected yield points observation due to random void distributions. It is worth noting that the characteristic dispersion of the yield surface is achieved such that there is little effect of the random distribution at low stress triaxiality, while the spread of the curves on either side of the mean increases with stress triaxiality until about $T\approx 4$. Moreover, the narrowing of the dispersion in the yield surface close to the mean stress axis is also captured (see Figs.~\ref{fig:YS_clusters_FCC_mean},~\ref{fig:STD}, and~\ref{fig:GursonS}). 

The proposed distribution-enriched GTN yield surface in Eq.~\eqref{eq:GursonS} can be employed in large-scale continuum modeling by assigning individual finite elements (or Gauss points) a new material value $\alpha$, which determines the yield surface for this particular material point. That is, the distribution-enriched yield surface may be expressed as
\begin{equation}
\label{eq:Gurson_alpha}
    \Phi = \frac{\sigma_e^2}{\sigma_0^2}+2q_1f \text{cosh}\left[\frac{3}{2}q_2\frac{\sigma_m}{\sigma_0}\right]-\left(1\pm \frac{\alpha}{\sigma_0} \right)^2\left(1+(q_1f)^2\right),
\end{equation}
where the $\alpha$-value could be assigned with a random spatial distribution throughout the volume such that it follows a normal distribution of the form
%
%\begin{equation}
%\label{eq:GursonS}
%    \psi(x) = \frac{1}{\sqrt{2\pi\textbf{S}^2}}\text{e}^{-\frac{(x-\mu)^2}{2\textbf{S}^2}}
%\end{equation}
%
%
%
\begin{equation}
\label{eq:alpha_dist}
    \psi(x) = \frac{1}{\sqrt{2\pi\textbf{S}^2}}\exp\left[{-\frac{((\alpha+\mu)-\mu)^2}{2\textbf{S}^2}}\right].
    %\psi(x) = \frac{1}{\sqrt{2\pi\textbf{S}^2}}\text{e}^{-\frac{\alpha^2}{2\textbf{S}^2}}
\end{equation}
Here, $\mu$ is the mean of the yield surface based on the unit cell calculations with random void distributions (see Fig.~\ref{fig:mean_Gurson}), and $\textbf{S}$ is the corresponding standard deviation (see Fig.~\ref{fig:STD-rho}). It would be expected that the interval $\alpha\in[\pm2\textbf{S}]$ would represent about $95\%$ of the yield point observations obtained from unit cell calculations, while $70\%$ of the observations would lie in the interval $\alpha\in[\pm\textbf{S}]$ (depicted in Fig.~\ref{fig:STD}). Adopting this procedure and introducing the yield surface from Eq.~\eqref{eq:Gurson_alpha} into a continuum-based finite element calculation would reflect the dispersion of yield points when using a discretization where individual integration points represent about four voids. Increasing the number of voids described by a yield surface in an integration point would need appropriate scaling of the standard deviation up until the limit, where each integration point describes a very large ensemble of voids and a mean yield surface is appropriate.
In this way, the resolution of the discretization will control the effect of the random void distributions.

\section{Conclusions}
\label{S:6}
\noindent{}The present work demonstrates how the dispersion of yield points, identified as the limit load, in ductile metals with random void distributions ties to the deformation mechanism and suggests a way to incorporate the findings into the Gurson-Tvergaard-Needleman yield surface. The study relies on a numerical investigation of a periodic microstructure represented by unit cells containing a limited number of randomly distributed voids. The unit cell setup is considered a mesoscale model of the material and it provided insight into the statistical characteristics of the yield locus owing to the void distribution. The key findings for the dispersion of the yield points are
\begin{itemize}
    \item \underline{A strong dependency on the stress state exists}. The dispersion due to random void distributions is practically zero at low triaxiality, while it grows to the largest value in the range of $4<T<5$ and drops slightly for higher triaxialities (see Figs.~\ref{fig:YS_clusters_FCC_mean} and~\ref{fig:STD}). The reason is found in the deformation mechanism at play as macroscopic yielding prevails at low triaxiality~\citep[in line with][]{tekouglu2015localization}, while a complex mixture of localization modes can develop at higher triaxiality depending on the intervoid distance and location of the voids (see discussion in Section~\ref{sec:Gurson_enriched}).
    \item \underline{The standard deviation follows the deformation mechanism}. The plastic index introduced in Eq.~\eqref{eq:PI} is adopted to demonstrate a correlation between the dispersion and the prevailing deformation mechanism. The index equals one when plastic straining occurs in the entire unit cell at the limit load, while small index values signal localization in a portion of the unit cell volume. It is found that the index displays a large spread for different randomization when $T\gtrsim4$ and that the index standard deviation of the observations correlates with that of the dispersion of the yield points. Thus, it is concluded that the variations in the localization mechanism determine the dispersion of the yield points.
    \item{} \underline{The porosity influences the spread of the yield locus}. The variation in the standard deviation of both $\textbf{PI}$ and $\mu$ depends on the porosity as the intervoid ligament size diminishes with increasing void volume fraction. The variation is most significant for a large porosity such that the peak value of $\textbf{S}$ attains the highest level at $\rho\approx0.8$ (corresponding to $T\approx4.3$) but also the largest relative drop at higher triaxialities (see Fig.~\ref{fig:STD-rho}). Thus, the width of the yield locus up until $T\approx4.3$ and the following narrowing near the mean stress axis increases with porosity (see Figs.~\ref{fig:STD} and \ref{fig:GursonS}). 
\end{itemize}

The present work investigates statistical variations of yield surfaces for porous materials. It is shown that the classical GTN yield surface rather accurately models the mean yield surface from the unit cell calculations with random void distributions in the full range of positive stress triaxialities (see Fig.~\ref{fig:mean_Gurson}).
%This is despite the localization mechanisms changing with the imposed stress state, while the GTN model is based on localization within intervoid ligaments. 
Thus, the classical GTN model may be used as a backbone model in a distribution-enriched extension that accounts for the dispersion of the yield point. The GTN model may be enriched by scaling its size by a factor of $(1\pm\textbf{S}/\sigma_0)^2$. This provides a model in good agreement yield surface obtained from unit cell calculations (see Figs.~\ref{fig:YS_clusters_FCC_mean},~\ref{fig:STD}, and~\ref{fig:GursonS}). Finally, a procedure for implementing the new GTN model is proposed in Section~\ref{sec:Gurson_enriched}.

%
% Acknowledgement
%
%
\section{Acknowledgements}
\noindent{}G. Papazafeiropoulos (NTUA) is acknowledged for his help in developing the void generation algorithm. This research was financially supported by Danish Council for Independent Research through the research project ``Advanced Damage Models with InTrinsic Size Effects'' (Grant no: DFF-7017-00121). Emilio Mart\'{\i}nez-Pa\~neda was supported by an UKRI Future Leaders Fellowship [grant MR/V024124/1].

%% References
%\def\urlprefix{doi: }
%\section*{References}
\bibliographystyle{elsart-harv}
\bibliography{sample.bib}

\begin{thebibliography}{26}
\expandafter\ifx\csname natexlab\endcsname\relax\def\natexlab#1{#1}\fi
\providecommand{\url}[1]{\texttt{#1}}
\providecommand{\href}[2]{#2}
\providecommand{\path}[1]{#1}
\providecommand{\DOIprefix}{doi:}
\providecommand{\ArXivprefix}{arXiv:}
\providecommand{\URLprefix}{URL: }
\providecommand{\Pubmedprefix}{pmid:}
\providecommand{\doi}[1]{\href{http://dx.doi.org/#1}{\path{#1}}}
\providecommand{\Pubmed}[1]{\href{pmid:#1}{\path{#1}}}
\providecommand{\bibinfo}[2]{#2}
\ifx\xfnm\relax \def\xfnm[#1]{\unskip,\space#1}\fi
%Type = Book
\bibitem[{Abaqus(2020)}]{ABAQUS_2016}
\bibinfo{author}{Abaqus}, \bibinfo{year}{2020}.
\newblock \bibinfo{title}{SIMULIA User Assistance 2020, Abaqus Documentation}.
\newblock \bibinfo{publisher}{{D}assault {S}yst\`emes {S}imulia {C}orp}.
%Type = Article
\bibitem[{Andersen et~al.(2020)Andersen, Tekoglu and
  Nielsen}]{Andersen_et_al_2020b}
\bibinfo{author}{Andersen, R.}, \bibinfo{author}{Tekoglu, C.},
  \bibinfo{author}{Nielsen, K.L.}, \bibinfo{year}{2020}.
\newblock \bibinfo{title}{{Cohesive traction-separation relations for tearing
  of ductile plates with randomly distributed void nucleation sites}}.
\newblock \bibinfo{journal}{Int. J. Fract.} \bibinfo{volume}{224},
  \bibinfo{pages}{187--198}.
\newblock \DOIprefix\doi{10.1007/s10704-020-00454-2}.
%Type = Article
\bibitem[{Becker(1987)}]{becker1987effect}
\bibinfo{author}{Becker, R.}, \bibinfo{year}{1987}.
\newblock \bibinfo{title}{The effect of porosity distribution on ductile
  failure}.
\newblock \bibinfo{journal}{Journal of the Mechanics and Physics of Solids}
  \bibinfo{volume}{35}, \bibinfo{pages}{577--599}.
\newblock \DOIprefix\doi{10.1016/0022-5096(87)90018-4}.
%Type = Article
\bibitem[{Benzerga and Leblond(2010)}]{Benzerga_Leblond_2010}
\bibinfo{author}{Benzerga, s.}, \bibinfo{author}{Leblond, J.B.},
  \bibinfo{year}{2010}.
\newblock \bibinfo{title}{Ductile fracture by void growth to coalescence}.
\newblock \bibinfo{journal}{Advances in Appl. Mech.} \bibinfo{volume}{44},
  \bibinfo{pages}{169--305}.
\newblock \DOIprefix\doi{10.1016/S0065-2156(10)44003-X}.
%Type = Article
\bibitem[{Buffiere et~al.(1999)Buffiere, Maire, Cloetens, Lormand and
  Fougeres}]{Buffiere_et_al_1999}
\bibinfo{author}{Buffiere, J.Y.}, \bibinfo{author}{Maire, E.},
  \bibinfo{author}{Cloetens, P.}, \bibinfo{author}{Lormand, G.},
  \bibinfo{author}{Fougeres, R.}, \bibinfo{year}{1999}.
\newblock \bibinfo{title}{Characterization of internal damage in a mmc$_p$
  using x-ray synchrotron phase contrast microtomography}.
\newblock \bibinfo{journal}{Acta Materialia} \bibinfo{volume}{47},
  \bibinfo{pages}{1613--1625}.
\newblock \DOIprefix\doi{10.1016/S1359-6454(99)00024-5}.
%Type = Article
\bibitem[{Cadet et~al.(2021)Cadet, Besson, Flouriot, Forest, Kerfriden and {de
  Rancourt}}]{Cadet2021}
\bibinfo{author}{Cadet, C.}, \bibinfo{author}{Besson, J.},
  \bibinfo{author}{Flouriot, S.}, \bibinfo{author}{Forest, S.},
  \bibinfo{author}{Kerfriden, P.}, \bibinfo{author}{{de Rancourt}, V.},
  \bibinfo{year}{2021}.
\newblock \bibinfo{title}{Strain localization analysis in materials containing
  randomly distributed voids: Competition between extension and shear failure
  modes}.
\newblock \bibinfo{journal}{International Journal of Fracture}
  \bibinfo{volume}{230}.
\newblock \DOIprefix\doi{10.1007/s10704-021-00562-7}.
%Type = Article
\bibitem[{Cadet et~al.(2022)Cadet, Besson, Flouriot, Forest, Kerfriden, Lacourt
  and {de Rancourt}}]{Cadet2022}
\bibinfo{author}{Cadet, C.}, \bibinfo{author}{Besson, J.},
  \bibinfo{author}{Flouriot, S.}, \bibinfo{author}{Forest, S.},
  \bibinfo{author}{Kerfriden, P.}, \bibinfo{author}{Lacourt, L.},
  \bibinfo{author}{{de Rancourt}, V.}, \bibinfo{year}{2022}.
\newblock \bibinfo{title}{Strain localization analysis in materials containing
  randomly distributed voids: Competition between extension and shear failure
  modes}.
\newblock \bibinfo{journal}{Journal of the Mechanics and Physics of Solids}
  \bibinfo{volume}{166}.
\newblock \DOIprefix\doi{10.1016/j.jmps.2022.104933}.
%Type = Article
\bibitem[{\c{C}elik et~al.(2021)\c{C}elik, Andersen, Teko\u{g}lu and
  Nielsen}]{Celik_et_al_2021}
\bibinfo{author}{\c{C}elik, {\c{S}}.}, \bibinfo{author}{Andersen, R.G.},
  \bibinfo{author}{Teko\u{g}lu, C.}, \bibinfo{author}{Nielsen, K.L.},
  \bibinfo{year}{2021}.
\newblock \bibinfo{title}{On the dependence of crack surface morphology and
  energy dissipation on microstructure in ductile plate tearing}.
\newblock \bibinfo{journal}{Int. J. Fract.}
  \DOIprefix\doi{10.1007/s10704-020-00513-8}.
%Type = Article
\bibitem[{Dubensky and Koss(1987)}]{dubensky1987void}
\bibinfo{author}{Dubensky, E.}, \bibinfo{author}{Koss, D.A.},
  \bibinfo{year}{1987}.
\newblock \bibinfo{title}{Void/pore distributions and ductile fracture}.
\newblock \bibinfo{journal}{Metallurgical Transactions A} \bibinfo{volume}{18},
  \bibinfo{pages}{1887--1895}.
\newblock \DOIprefix\doi{10.1007/BF02647018}.
%Type = Article
\bibitem[{Fritzen et~al.(2012)Fritzen, Forest, Böhlke, Kondo and
  Kanit}]{Fritzen2012}
\bibinfo{author}{Fritzen, F.}, \bibinfo{author}{Forest, S.},
  \bibinfo{author}{Böhlke, T.}, \bibinfo{author}{Kondo, D.},
  \bibinfo{author}{Kanit, T.}, \bibinfo{year}{2012}.
\newblock \bibinfo{title}{Computational homogenization of elasto-plastic porous
  metals}.
\newblock \bibinfo{journal}{International Journal of Plasticity}
  \bibinfo{volume}{29}, \bibinfo{pages}{102--119}.
\newblock \DOIprefix\doi{10.1016/j.ijplas.2011.08.005}.
%Type = Article
\bibitem[{Guo and Wong(2018)}]{guo2018void}
\bibinfo{author}{Guo, T.}, \bibinfo{author}{Wong, W.}, \bibinfo{year}{2018}.
\newblock \bibinfo{title}{Void-sheet analysis on macroscopic strain
  localization and void coalescence}.
\newblock \bibinfo{journal}{Journal of the Mechanics and Physics of Solids}
  \bibinfo{volume}{118}, \bibinfo{pages}{172--203}.
\newblock \DOIprefix\doi{10.1016/j.jmps.2018.05.002}.
%Type = Article
\bibitem[{Gurson(1977)}]{gurson1977}
\bibinfo{author}{Gurson, A.L.}, \bibinfo{year}{1977}.
\newblock \bibinfo{title}{Continuum theory of ductile rupture by void
  nucleation and growth. part {I}: yield criteria and flow rules for porous
  ductile media}.
\newblock \bibinfo{journal}{Journal of Engineering Materials and Technology}
  \bibinfo{volume}{99}, \bibinfo{pages}{2--15}.
\newblock \DOIprefix\doi{10.1115/1.3443401}.
%Type = Article
\bibitem[{Hannard et~al.(2017)Hannard, Castin, Maire, Mokso, Pardoen and
  Simar}]{hannard2017ductilization}
\bibinfo{author}{Hannard, F.}, \bibinfo{author}{Castin, S.},
  \bibinfo{author}{Maire, E.}, \bibinfo{author}{Mokso, R.},
  \bibinfo{author}{Pardoen, T.}, \bibinfo{author}{Simar, A.},
  \bibinfo{year}{2017}.
\newblock \bibinfo{title}{Ductilization of aluminium alloy 6056 by friction
  stir processing}.
\newblock \bibinfo{journal}{Acta Materialia} \bibinfo{volume}{130},
  \bibinfo{pages}{121--136}.
\newblock \DOIprefix\doi{10.1016/j.actamat.2017.01.047}.
%Type = Article
\bibitem[{Holte et~al.(2021)Holte, Srivastava, Mart{\'\i}nez-Pa{\~n}eda,
  Niordson and Nielsen}]{holte2021interaction}
\bibinfo{author}{Holte, I.}, \bibinfo{author}{Srivastava, A.},
  \bibinfo{author}{Mart{\'\i}nez-Pa{\~n}eda, E.}, \bibinfo{author}{Niordson,
  C.F.}, \bibinfo{author}{Nielsen, K.L.}, \bibinfo{year}{2021}.
\newblock \bibinfo{title}{Interaction of void spacing and material size effect
  on inter-void flow localization}.
\newblock \bibinfo{journal}{Journal of Applied Mechanics} \bibinfo{volume}{88},
  \bibinfo{pages}{021010}.
\newblock \DOIprefix\doi{10.1115/1.4049022}.
%Type = Article
\bibitem[{Hure(2021)}]{hure2021yield}
\bibinfo{author}{Hure, J.}, \bibinfo{year}{2021}.
\newblock \bibinfo{title}{Yield criterion and finite strain behavior of random
  porous isotropic materials}.
\newblock \bibinfo{journal}{European Journal of Mechanics-A/Solids}
  \bibinfo{volume}{85}, \bibinfo{pages}{104143}.
\newblock \DOIprefix\doi{10.1016/j.euromechsol.2020.104143}.
%Type = Article
\bibitem[{Khdir et~al.(2015)Khdir, Kanit, Zaïri and
  Naït-Abdelaziz}]{Khdir2015}
\bibinfo{author}{Khdir, Y.K.}, \bibinfo{author}{Kanit, T.},
  \bibinfo{author}{Zaïri, F.}, \bibinfo{author}{Naït-Abdelaziz, M.},
  \bibinfo{year}{2015}.
\newblock \bibinfo{title}{A computational homogenization of random porous
  media: Effect of void shape and void content on the overall yield surface}.
\newblock \bibinfo{journal}{European Journal of Mechanics - A/Solids}
  \bibinfo{volume}{49}, \bibinfo{pages}{137--145}.
\newblock \DOIprefix\doi{10.1016/j.euromechsol.2014.07.001}.
%Type = Article
\bibitem[{Lecarme et~al.(2014)Lecarme, Maire, Arun~Kumar, De~Vleeschouwer,
  Jacques, Simar and Pardoen}]{Lecarme_et_al_2014}
\bibinfo{author}{Lecarme, L.}, \bibinfo{author}{Maire, E.},
  \bibinfo{author}{Arun~Kumar, K.C.}, \bibinfo{author}{De~Vleeschouwer, C.},
  \bibinfo{author}{Jacques, L.}, \bibinfo{author}{Simar, A.},
  \bibinfo{author}{Pardoen, T.}, \bibinfo{year}{2014}.
\newblock \bibinfo{title}{Heterogeneous void growth revealed by in situ 3-d
  x-ray microtomography using automatic cavity tracking}.
\newblock \bibinfo{journal}{Acta Materialia} \bibinfo{volume}{63},
  \bibinfo{pages}{130--139}.
\newblock \DOIprefix\doi{10.1016/j.actamat.2013.10.014}.
%Type = Article
\bibitem[{Liu et~al.(2019)Liu, Zheng, Osovski and
  Srivastava}]{liu2019micromechanism}
\bibinfo{author}{Liu, Y.}, \bibinfo{author}{Zheng, X.},
  \bibinfo{author}{Osovski, S.}, \bibinfo{author}{Srivastava, A.},
  \bibinfo{year}{2019}.
\newblock \bibinfo{title}{On the micromechanism of inclusion driven ductile
  fracture and its implications on fracture toughness}.
\newblock \bibinfo{journal}{Journal of the Mechanics and Physics of Solids}
  \bibinfo{volume}{130}, \bibinfo{pages}{21--34}.
\newblock \DOIprefix\doi{10.1016/j.jmps.2019.05.010}.
%Type = Article
\bibitem[{Magnusen et~al.(1988)Magnusen, Dubensky and
  Koss}]{magnusen1988effect}
\bibinfo{author}{Magnusen, P.}, \bibinfo{author}{Dubensky, E.},
  \bibinfo{author}{Koss, D.}, \bibinfo{year}{1988}.
\newblock \bibinfo{title}{The effect of void arrays on void linking during
  ductile fracture}.
\newblock \bibinfo{journal}{Acta Metallurgica} \bibinfo{volume}{36},
  \bibinfo{pages}{1503--1509}.
\newblock \DOIprefix\doi{10.1016/0001-6160(88)90217-9}.
%Type = Article
\bibitem[{Papazafeiropoulos et~al.(2017)Papazafeiropoulos, Mu{\~{n}}iz-Calvente
  and Mart{\'{i}}nez-Pa{\~{n}}eda}]{AES2017}
\bibinfo{author}{Papazafeiropoulos, G.}, \bibinfo{author}{Mu{\~{n}}iz-Calvente,
  M.}, \bibinfo{author}{Mart{\'{i}}nez-Pa{\~{n}}eda, E.}, \bibinfo{year}{2017}.
\newblock \bibinfo{title}{{Abaqus2Matlab: A suitable tool for finite element
  post-processing}}.
\newblock \bibinfo{journal}{Advances in Engineering Software}
  \bibinfo{volume}{105}, \bibinfo{pages}{9--16}.
\newblock \DOIprefix\doi{10.1016/j.advengsoft.2017.01.006}.
%Type = Article
\bibitem[{Perrin and Leblond(1990)}]{perrin1990analytical}
\bibinfo{author}{Perrin, G.}, \bibinfo{author}{Leblond, J.B.},
  \bibinfo{year}{1990}.
\newblock \bibinfo{title}{Analytical study of a hollow sphere made of plastic
  porous material and subjected to hydrostatic tension-application to some
  problems in ductile fracture of metals}.
\newblock \bibinfo{journal}{International Journal of Plasticity}
  \bibinfo{volume}{6}, \bibinfo{pages}{677--699}.
\newblock \DOIprefix\doi{10.1016/0749-6419(90)90039-H}.
%Type = Article
\bibitem[{Teko\~glu and Nielsen(2019)}]{Tekoglu_Nielsen_2019}
\bibinfo{author}{Teko\~glu, C.}, \bibinfo{author}{Nielsen, K.L.},
  \bibinfo{year}{2019}.
\newblock \bibinfo{title}{{Effect of damage-related microstructural parameters
  on plate tearing at steady state}}.
\newblock \bibinfo{journal}{Eur. J. Mech. A Solids A/Solids}
  \bibinfo{volume}{77}, \bibinfo{pages}{103818 \\*}.
\newblock \DOIprefix\doi{10.1016/j.euromechsol.2019.103818}.
%Type = Article
\bibitem[{Teko{\u{g}}lu et~al.(2015)Teko{\u{g}}lu, Hutchinson and
  Pardoen}]{tekouglu2015localization}
\bibinfo{author}{Teko{\u{g}}lu, C.}, \bibinfo{author}{Hutchinson, J.},
  \bibinfo{author}{Pardoen, T.}, \bibinfo{year}{2015}.
\newblock \bibinfo{title}{On localization and void coalescence as a precursor
  to ductile fracture}.
\newblock \bibinfo{journal}{Philosophical Transactions of the Royal Society A:
  Mathematical, Physical and Engineering Sciences} \bibinfo{volume}{373},
  \bibinfo{pages}{20140121}.
\newblock \DOIprefix\doi{10.1098/rsta.2014.0121}.
%Type = Article
\bibitem[{Tvergaard(1981)}]{tvergaard1981}
\bibinfo{author}{Tvergaard, V.}, \bibinfo{year}{1981}.
\newblock \bibinfo{title}{Influence of voids on shear band instabilities under
  plane strain conditions}.
\newblock \bibinfo{journal}{International Journal of Fracture}
  \bibinfo{volume}{17}, \bibinfo{pages}{389--407}.
\newblock \DOIprefix\doi{10.1007/BF00036191}.
%Type = Article
\bibitem[{Tvergaard(1990)}]{Tvergaard_1990}
\bibinfo{author}{Tvergaard, V.}, \bibinfo{year}{1990}.
\newblock \bibinfo{title}{{Material failure by void growth to coalescence}}.
\newblock \bibinfo{journal}{Adv. Appl. Mech.} \bibinfo{volume}{27},
  \bibinfo{pages}{83--151}.
\newblock \DOIprefix\doi{10.1016/S0065-2156(08)70195-9}.
%Type = Article
\bibitem[{Tvergaard and Needleman(1984)}]{tvergaard1984analysis}
\bibinfo{author}{Tvergaard, V.}, \bibinfo{author}{Needleman, A.},
  \bibinfo{year}{1984}.
\newblock \bibinfo{title}{Analysis of the cup-cone fracture in a round tensile
  bar}.
\newblock \bibinfo{journal}{Acta metallurgica} \bibinfo{volume}{32},
  \bibinfo{pages}{157--169}.
\newblock \DOIprefix\doi{10.1016/0001-6160(84)90213-X}.

\end{thebibliography}
%
%
%\linespread{1}
%\section*{Tables}
%\input{Table}
%
\clearpage
\newpage
\linespread{1}
\section*{Figures}
% Figures
%
\renewcommand{\thefigure}{\arabic{figure}}    
\setcounter{figure}{0}  % reset counter 
%
%
%
%---------------------------------------------------
% Fig 1
\begin{figure}[h]
\centering
\includegraphics[width=0.5\textwidth]{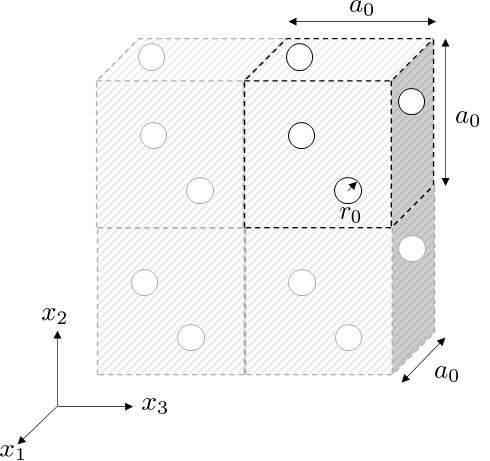}
\caption{Schematic of the periodic arrangement of the mesoscale unit cell containing a random distribution of four equal-sized spherical voids with initial radius $r_0$. The unit cell is repeated along all coordinate axis and has side lengths $a_0$ in all directions.}
    \label{fig:unitcellmodel}
\end{figure}
%---------------------------------------------------

%---------------------------------------------------
% Fig 2
\begin{figure}[h]
\centering
\includegraphics[width=0.95\textwidth]{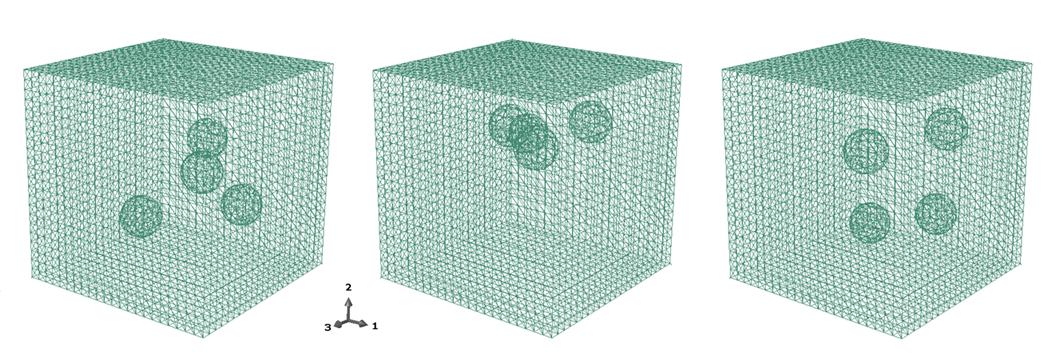}
\caption{Examples of three unit cell finite element meshes containing different randomizations of voids.}
    \label{fig:meshes}
\end{figure}
%---------------------------------------------------

%---------------------------------------------------
% Fig 3
\begin{figure}[h]
\centering
\subfigure[]
{
\includegraphics[width=0.7\textwidth]{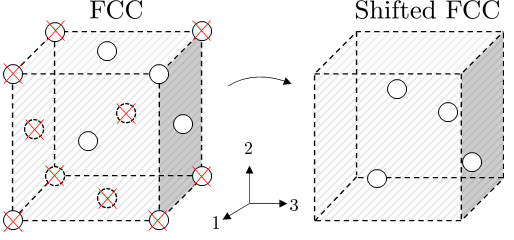}
}
\subfigure[]
{
\includegraphics[width=0.35\textwidth]{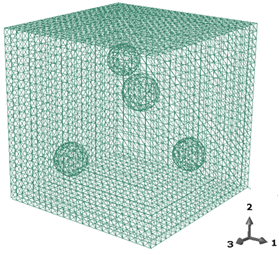}
}
\vspace{-0.35cm}
\caption{Procedure for modeling a Face-Centered Configuration (FCC) of the voided unit cell, showing (a) the FCC unit cell where the voids removed (left) are outside the shifted unit cell (right), and (b) the corresponding finite element mesh of the FCC unit cell.}
    \label{fig:FCC-shifted}
\end{figure}
%---------------------------------------------------

%---------------------------------------------------
% Fig 4
\begin{figure}[h]
\centering
    \includegraphics[width=0.55\textwidth]{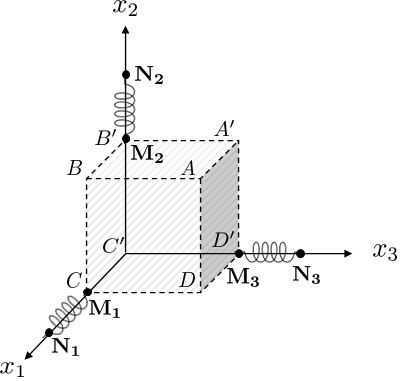}
    \vspace{-0.35cm}
\caption{Modeling procedure for imposing periodic boundary conditions and loading to the unit cell, showing the faces, edges, and vertices defined by $A, B, C, D$ and $A', B', C, D'$. The dummy nodes to control the imposed stress state are denoted $N_i$, and the related master nodes are $M_i$ $(i=1,2,3)$, which are part of the finite element mesh. The dummy and master nodes are connected with spring elements, as illustrated.}
\label{fig:RVE_PBC}
\end{figure}
%---------------------------------------------------

%---------------------------------------------------
% Fig 5
\begin{figure}[h]
\centering
\includegraphics[width=0.7\textwidth]{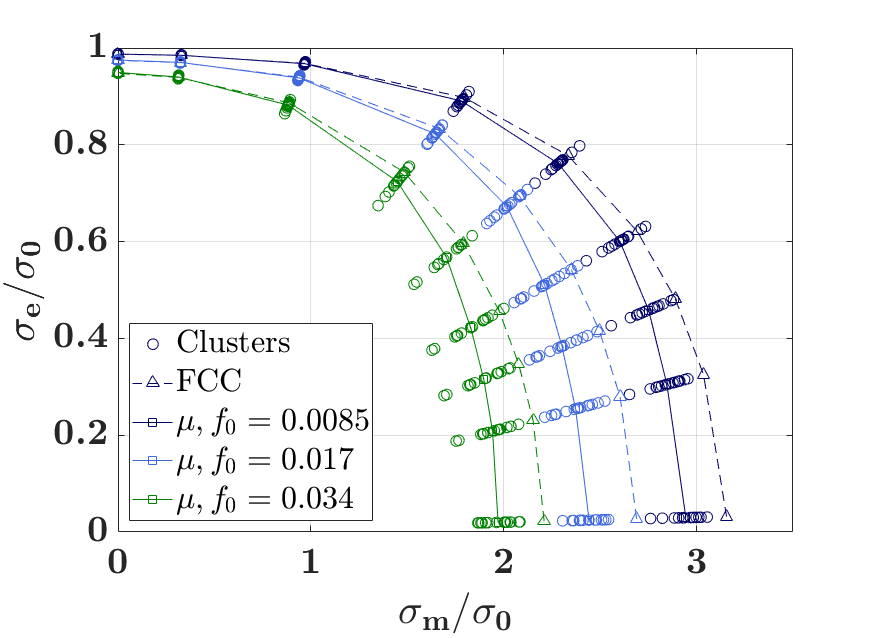}
\caption[]{Yield points for different combinations of porosity $f_0$, stress state $T$, and void distribution (15 different randomizations of each $f_0$-$T$-combination), alongside the calculated mean values (solid lines) and results for the FCC unit cell (dashed line). The stress state is limited to axisymmetry and controlled by the $\rho=\sigma_{2}/\sigma_{1}=\sigma_{3}/\sigma_{1}$ as outlined in Section~\ref{sec:numericalmethod}. Colors distinguish the different porosity values.}
%
%\caption[]{The 15 different clusters plotted along with the FCC distribution for all three levels of porosity, $f_0=0.0085, 0,017$ and $0.34$, and loading conditions, with $\rho$-values from -0.5 to 1. The mean yield surface given as the mean distance from origin calculated based on the 15 clusters (not FCC) is also shown for the three values of $f_0$ considered.}
\label{fig:YS_clusters_FCC_mean}
\end{figure} 
%---------------------------------------------------

%---------------------------------------------------
% Fig 6
\begin{figure}[h]
\centering
\includegraphics[width=0.7\textwidth]{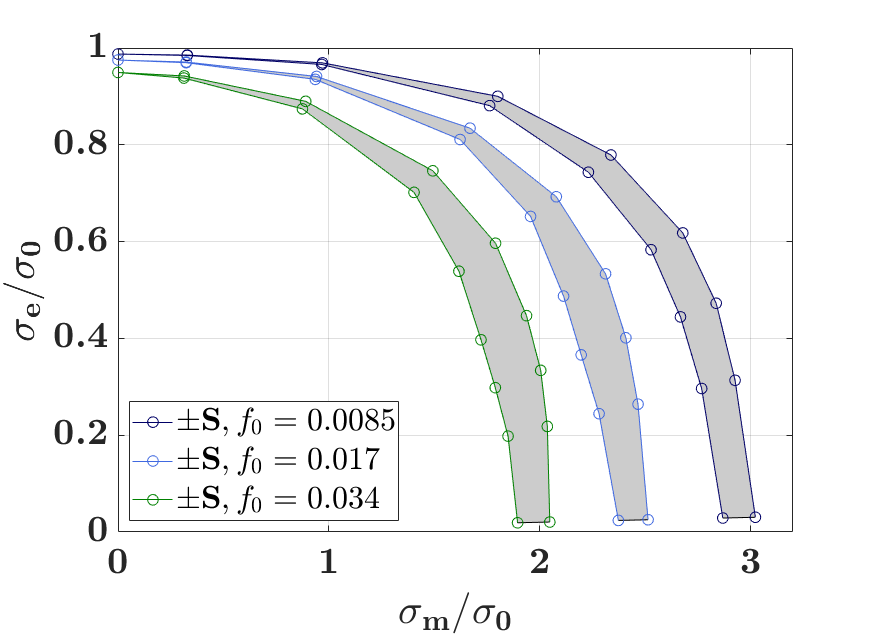}
\caption[]{Yield surfaces constructed from the unit cell results based on the mean distance $\mu$ from the origin in the von Mises versus mean stress space, and the corresponding standard deviation $\textbf{S}/\sigma_0$ (see Section~\ref{S:4}). The grey-shaded region encloses about $70\%$ of the yield points. Colors distinguish the different porosity values.}    
%
%\caption[]{Range of standard deviation of mean distance from origin for the three values of $f_0$ considered.}
\label{fig:STD}
\end{figure} 
%---------------------------------------------------

%---------------------------------------------------
% Fig 7
\begin{figure}[h]
\centering
    \includegraphics[width=0.7\textwidth]{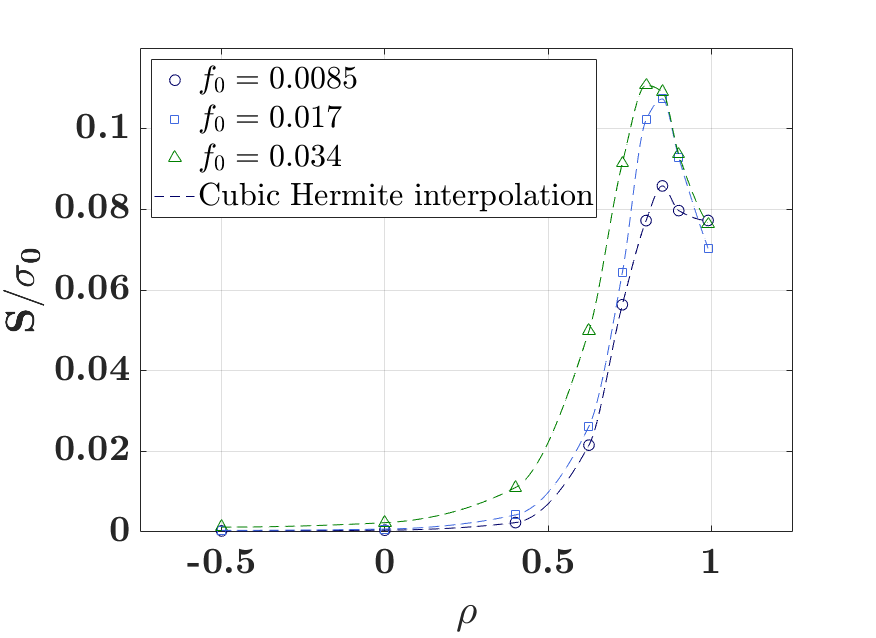}
\caption[]{Standard deviation $\mathbf{S}/\sigma_0$ of the mean distance from the origin to the yield surface normalized by the yield stress $\sigma_0$ as a function of the applied stress ratio $\rho$ for the three porosity-values $f_0$ considered. The dashed curves represent a piece-wise cubic Hermite interpolation of the discrete data points. Colors distinguish the different porosity values.}
\label{fig:STD-rho}
\end{figure} 
%---------------------------------------------------

%---------------------------------------------------
% Fig 8
\begin{figure}[h]
\centering
    \includegraphics[width=0.7\textwidth]{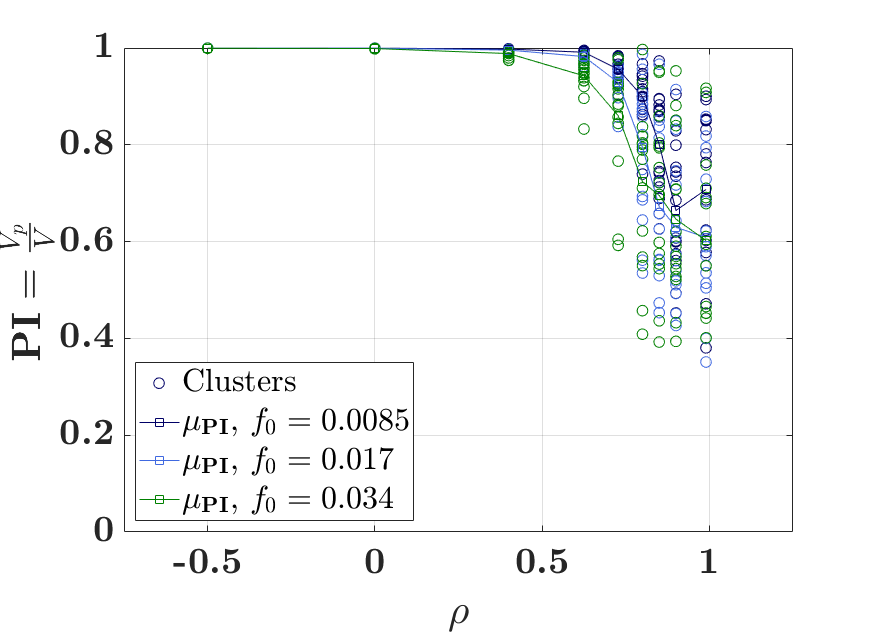}
\caption[]{The plastic index $\textbf{PI}$ introduced in Eq.\eqref{eq:PI} ~for all combinations of porosity $f_0$, stress state $T$, and void distributions (15 different randomizations of each $f_0$-$T$-combination) as a function of the ratio $\rho$. The mean plastic index $\mu_{\mathbf{PI}}$ is also shown as solid lines. Colors distinguish the different porosity values.}
\label{fig:PI}
\end{figure} 
%---------------------------------------------------

%---------------------------------------------------
% Fig 9
\begin{figure}[h]
\centering
    \includegraphics[width=0.7\textwidth]{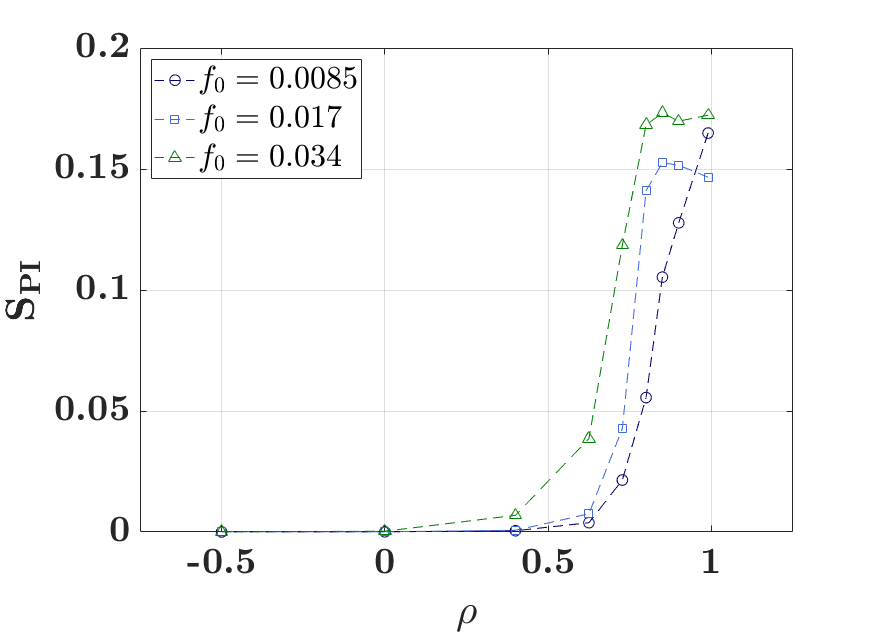}
%\caption[]{Standard deviation of the plastic index, $\mathbf{S_{PI}}$, as function of the applied $\rho$ for all values of $f_0$ considered.}
%
\caption{Standard deviation $\mathbf{S_{PI}}$ of the plastic index $\mathbf{PI}$ (see Eq.~\ref{eq:PI}) as a function of the applied stress ratio $\rho$ for the three porosity-values $f_0$ considered. The dashed curves represent a piece-wise cubic Hermite interpolation of the discrete data points. Colors distinguish the different porosity values.}
\label{fig:std_PI}
\end{figure}

%---------------------------------------------------
% Fig 10
\begin{figure}[h]
\centering
    \includegraphics[width=0.7\textwidth]{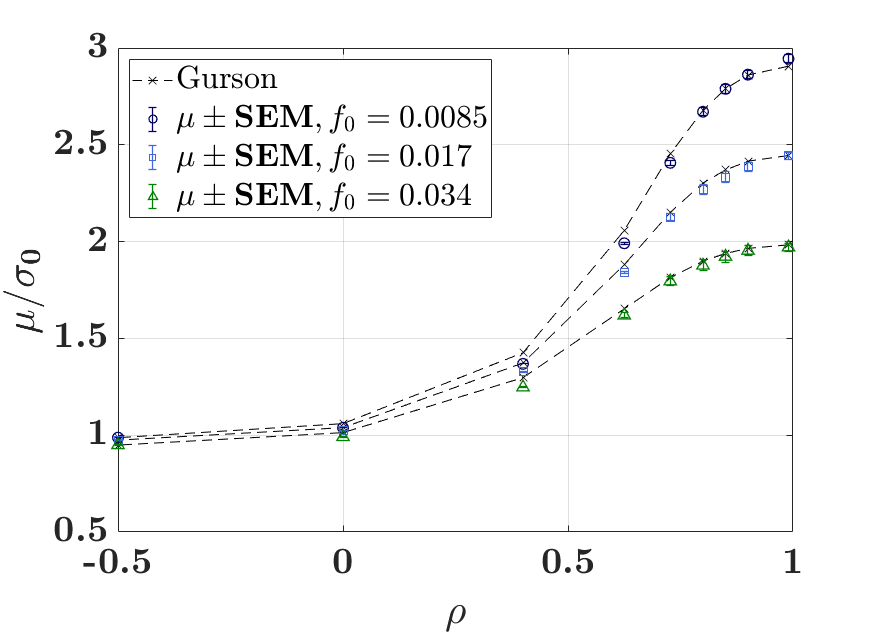}
\caption[]{Mean distance from origin $\mu$ form the unit cell calculations as a function of the stress ratio $\rho$, alongside with error bars indicating the standard error of mean $\mathbf{SEM}=\mathbf{S}/\sqrt{n}$. Here is $\mathbf{S}$ the standard deviation, and $n$ is the sample size ($n=15$). The results are shown together with the predictions for the classical Gurson-Tvergaard-Needleman (GTN) model for the three values of $f_0$ considered. Colors distinguish the different porosity values.}
\label{fig:mean_Gurson}
\end{figure} 
%---------------------------------------------------

%---------------------------------------------------
% Fig 11
\begin{figure}[h]
\centering
    \includegraphics[width=0.7\textwidth]{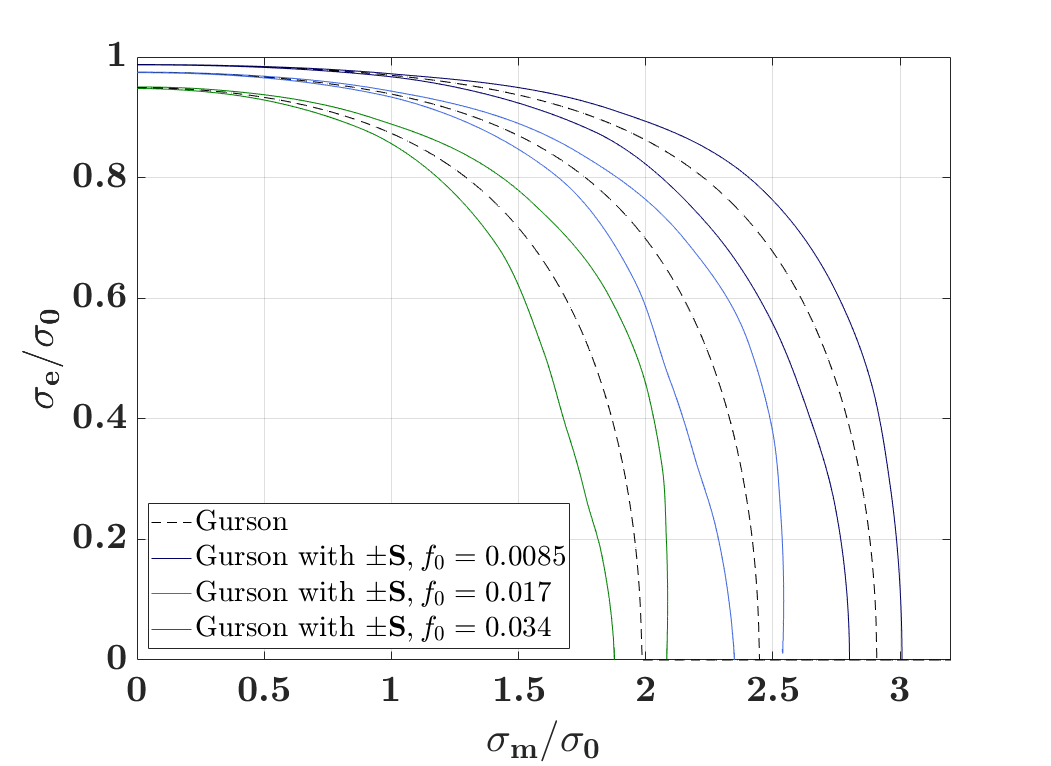}
\caption[]{Yield surfaces constructed from the new distribution-enriched Gurson-Tvergaard-Needleman (GTN) model in Eq.~\eqref{eq:GursonS}, showing the mean curve as a dashed line (repressed by the classical GTN model) and the dispersion of the yield surface represented by a confidence interval of $\pm\textbf{S}/\sigma_0$ (enclosing $70\%$ of the yield point observations). The yield surfaces are illustrated for the three initial porosity values considered distinguished by different colors.}
\label{fig:GursonS}
\end{figure} 
% \end{thebibliography}
%
\end{document}